 \documentclass[final,5p,times,twocolumn,numbers]{elsarticle}

\usepackage{amssymb}
\usepackage{lipsum}
\usepackage{physics}
\usepackage{subcaption}
\usepackage{graphicx}

\usepackage{xcolor}
\usepackage{comment}

\journal{Physics Letters B}

\begin{document}

\begin{frontmatter}



\title{Scalar diquark  mass and quark--diquark potential  from lattice QCD using
  the potential method with a static quark}


\author[first]{Kai-Wen Kelvin-Lee}
\ead{kelvin@rcnp.osaka-u.ac.jp}
\author[first]{Noriyoshi Ishii}
\ead{ishiin@rcnp.osaka-u.ac.jp}
\affiliation[first]{organization={Research Center for Nuclear Physics (RCNP),
The University of Osaka},
            addressline={10-1, Mihogaoka}, city={Ibaraki-shi},
            postcode={567-0047}, state={Osaka}, country={Japan}}

\begin{abstract}
We study the  scalar diquark mass and the  quark--diquark potential by applying
a HAL  QCD-inspired potential  method to  a baryonic  system composed of a
scalar diquark and a static quark.
The diquark mass is determined self-consistently by requiring that the
P-wave baryonic  spectrum  obtained  from  two-point
correlators  be reproduced within the potential framework.
Numerical  calculations are  performed  using $2+1$  flavor QCD  gauge
configurations  generated by  the  PACS-CS Collaboration  on a  $L^{3} \times T
=  32^{3} \times 64$ lattice with  $a^{-1} \approx 2.176$~GeV and the pion mass,
$m_{\pi} \approx 702$ MeV.
From the analysis,  we obtain a scalar diquark mass  which is close to the
na\"{\i}ve constituent quark estimate $ (2/3)m_{N}$, together with a
quark--diquark potential of the Cornell type (Coulomb + linear).
%
The string tension extracted  from the quark--diquark potential agrees within
approximately   5\%  with   that  obtained  from   the  static quark--antiquark
potential (Wilson Loop).
\end{abstract}



\begin{keyword}
lattice QCD \sep hadron structure \sep diquark mass \sep quark-diquark potential
\sep static quark



\end{keyword}

\end{frontmatter}




\section{Introduction}
\label{introduction}
Understanding the internal structure of hadrons remains a central issue in
quantum chromodynamics (QCD). 
Among the possible effective degrees of freedom inside baryons,
diquarks—correlated pairs of quarks—have long been proposed as important
building blocks of hadronic systems~\cite{Jaffe_2005}.
The color decomposition of two quarks,
\(
\mathbf{3}_c \otimes \mathbf{3}_c
=
\bar{\mathbf{3}}_c
\oplus
\mathbf{6}_c ,
\)
implies that diquarks necessarily carry non-vanishing color charge and therefore
cannot exist as isolated physical states.

Particular attention has been devoted to the scalar diquark with quantum numbers
\(J^P=0^+\), isospin \(I=0\), and color representation \(\bar{\mathbf{3}}_c\).
This configuration, often referred to as the ``good'' diquark, is expected to be
energetically favored.  
Model studies attribute its stability to attractive color-magnetic interactions
in constituent quark models as well as to instanton-induced forces in
QCD~\cite{tHooft_1976,Shuryak_1982,Schäfer_Shuryak_1998,Shuryak_Zahed_2004}.
Furthermore, at sufficiently high baryon density, scalar diquarks are predicted
to serve as Cooper pairs, leading to the color-superconducting phase of dense
QCD matter.

A   quantitative   understanding   of  diquark   properties   requires
nonperturbative calculations based on lattice QCD.
However,  the colored  nature  of diquarks  makes their  investigation
considerably more complicated than that of ordinary hadrons.
Because isolated  colored objects cannot appear  as asymptotic states, the
existence of  physical bound-state  poles analogous  to those  of ordinary
hadrons cannot be assumed for their correlation functions
(see Sec.~60.6.3  of Ref.~\cite{Workman_2022}).
This consideration applies equally to diquarks.
As a consequence, the standard procedure used for hadrons---extracting masses
from single-exponential  fits to  two-point correlators---does not have a clear
theoretical justification in this case.

Despite  this  conceptual  difficulty, several  strategies  have  been developed
in lattice QCD to probe diquark masses.
One commonly adopted approach is  to evaluate diquark correlators in a fixed
gauge,  typically the Landau  gauge, and extract  diquark masses from the
resulting temporal  correlator~\cite{Hess_1997, Babich_2007, Bi_2016}.
%
%
Although  plateaus are  observed in  the corresponding  effective mass plots,
the physical  interpretation of  the extracted  masses remains subtle,
%
because the existence  of a physical pole associated  with an isolated colored
object cannot be assumed.

Another line  of investigation  constructs gauge-invariant  systems by coupling
the diquark to a static color
source~\cite{C_Alexandrou_2006,A_Francis_2022,Orginos_2006,Green_2010}.
In this framework, baryon-like correlators  composed of a static quark and a
diquark are analyzed, and information on diquark mass splittings is    inferred
from    the    energy    differences   among    these
states~\cite{C_Alexandrou_2006,A_Francis_2022,Orginos_2006}.
%
While gauge invariance  is manifest in this  formulation,
the  measured energies  contain  contributions both  from the  diquark masses
and the quark-diquark interaction energies.
Therefore,  energy  splittings  among baryon-like  states  cannot,  in general,
be interpreted solely in terms of diquark mass splittings.

More recently, a different strategy has been proposed by extending the HAL QCD
method to quark--diquark systems, where the diquark is treated as an effective
degree of freedom  interacting with a quark through an effective nonrelativistic
potential~\cite{watanabe_2021, K_watanabe_2022, Nishioka_2025,
Nishioka:2026oca}.
%
In this  quark--diquark potential framework, the  diquark mass appears as a
parameter in  the quark--diquark  Hamiltonian and  is determined
self-consistently   together   with  the   effective   nonrelativistic
potential.
Applications of this  method to charmed baryons  such as \(\Lambda_c\) and
\(\Sigma_c\)  have demonstrated that  it provides a viable  way to circumvent
the absence of bound-state poles in diquark correlators.
%
However, the extracted diquark mass depends on the choice of the charm quark
mass   used  in  the  effective   Hamiltonian,  introducing  an additional
source of systematic uncertainty.

In the  present work we develop  an improved framework that  can avoid this
systematic  uncertainty by  replacing  the  charm quark  in  the quark–diquark
system with a static color source.
%
Within  the  quark--diquark  potential  framework,  this  modification allows us
to determine  the quark–diquark interaction potential while simultaneously
extracting  the  diquark  mass without  relying  on  a specific choice of the
charm quark mass.
%
Moreover, the  method provides a  framework in which the  diquark mass can be
disentangled from other  contributions in the total  energy of the  baryon-like
system,  thereby reducing  systematic  uncertainties associated with previous
determinations of diquark masses.

Using this approach, we perform a lattice QCD study of the scalar ``good''
diquark.
We determine its mass and extract the corresponding quark–diquark potential,
providing new quantitative insight into the dynamics of diquark correlations in
QCD.

\section{Theoretical Framework}

To circumvent  the difficulty of  extracting the diquark  mass without relying
on  the existence  of the  bound state  pole in  the two-point diquark
correlators, we adopt the strategy similar to that employed in
Ref.~\cite{watanabe_2021,        K_watanabe_2022,       Nishioka_2025,
Nishioka:2026oca}.
In particular, we  replace the charm quark used in  these studies by a static
quark of infinite mass.
Our aim  is to remove  the systematic uncertainty associated  with the choice of
the charm quark mass.
%

In the present setup, the quantum numbers of the baryonic states are essentially
determined by the orbital angular momentum $L$ of the diquark relative to the
static quark.
This simplification arises for two reasons. 
First, our analysis focuses exclusively on the scalar diquark channel. Second,
the spin degree of freedom of the static quark completely decouples from the
rest of the system.
This follows from heavy-quark spin symmetry, since the static quark corresponds
to the infinite-mass limit of a heavy quark.
Therefore, we label  the baryonic states as  $|B(L)\rangle$, where $L$ denotes
the relative orbital angular  momentum between the diquark and the static quark.

We introduce the  equal-time Nambu–Bethe–Salpeter (NBS) wavefunction for the
baryonic system composed of  a diquark and a  static quark as
\begin{equation}
  \psi_{L} (\mathbf{r})
  \equiv
  \bra{0}
  D_{c}(\mathbf{x})
  Q_{c}(\mathbf{y})
  \ket{B(L)}, 
  \quad
  \mathbf{r}
  \equiv
  \mathbf{x} - \mathbf{y},
\end{equation}
where  $Q_{c}(x)$ is  the static  quark field,  and $D_{c}(x)$  is the composite
scalar diquark field defined as
\begin{equation}
  D_{c}(x)
  \equiv
  \epsilon_{abc}
  u^{T}_{a}(x)
  C\gamma_{5}
  d_{b}(x),
\end{equation}
with $C \equiv i\gamma^2\gamma^0$ being the charge conjugation matrix and
$a,b,c$ denote color indices.

We require that the NBS wavefunction satisfy the following Schrödinger equation:
\begin{equation}
  \left(
  -\frac{\nabla^2}{2m_{D}}
  + \hat{V}
  \right)
  \psi_{L} (\mathbf{r})
  =
  (\varepsilon_{L} - m_{D})
  \psi_{L} (\mathbf{r}),
  \label{eq:schroedinger-eq}
\end{equation}
where  $\varepsilon_{L}$  is  the  total relativistic  energy  of  the baryonic
state  $\ket{B(L)}$, and  the difference  $(\varepsilon_{L} - m_{D})$ represents
the ``binding energy'' of the system.
The parameter  $m_{D}$ denotes the  diquark mass,
which  is determined  later by  matching  the baryonic  masses in  the P-wave
sector     obtained      by     Schr\"odinger     equation,
Eq.~(\ref{eq:schroedinger-eq}),   with   those  extracted   from   the
corresponding two-point correlation functions.
The  quark–diquark potential  operator $\hat{V}$  can be  expanded in terms  of
derivatives  as $\hat{V}  \simeq V_{0}(r)  + O(\nabla^{2})$, where $V_{0}(r)$ is
the central, spin-independent potential.
Because we consider a scalar  diquark ($J^{P} = 0^+$), both spin–spin and tensor
interactions  are absent.
Furthermore, the spin–orbit interactions vanish identically.
This is due to  the scalar nature of the diquark  ($J^P = 0^+$), which possesses
no spin, combined with the  fact that the spin of the static quark  completely
decouples  from the  dynamics of  the system  in the heavy-quark limit.

The equal-time  NBS wavefunction  is extracted from  the quark–diquark
four-point correlator.
For  $t >  t_{\rm src}$,  our quark-diquark  four-point correlator  is arranged
as
\begin{align}
  &C(\mathbf{r}, t; t_{\rm src}) \\
  &\equiv
  \frac{1}{V} \sum_{\mathbf{\Delta}} 
  \bra{0}
  D_c(\mathbf{r} + \mathbf{\Delta}, t) Q_c(\mathbf{\Delta}, t) \cdot \mathcal{J}^\dagger(t_{\rm src})
  \ket{0} \nonumber \\
  &= \frac{1}{V} \sum_{\mathbf{\Delta}} \sum_{n} 
  \bra{0} D_c(\mathbf{r} + \mathbf{\Delta}, t) Q_c(\mathbf{\Delta}, t) \ket{n} 
  \bra{n} \mathcal{J}^\dagger(t_{\rm src}) \ket{0} \nonumber \cdot e^{-E_{n}(t - t_{\rm src})}, \nonumber
\end{align}
where $V$ is the three dimensional spatial volume and $\Delta$ denotes a spatial
translation vector.
The  average  over  $\Delta$  is   taken  to  make  the  translational
invariance manifest.
%
%
The  states   $\ket{n}$  and   energies  $E_{n}$  denote   the  $n$-th
eigenstates and eigenenergies of the QCD Hamiltonian, respectively.
$\mathcal{J}(t_{\rm  src})$ denotes the source operator.
We explicitly specify the source  here. First, define $q_a(f,t) \equiv
\sum_{\mathbf{x}} q_a(\mathbf{x},t)  f(\mathbf{x})$ for $q =  u, d, Q$ with
$f(\mathbf{x}): \mathbb{R}^3 \to \mathbb{C}$.
Then, $D_c(f,t) \equiv  \epsilon_{abc} u_a^T(f,t) C\gamma_5 d_b(f,t)$, and the
source operator is written as
\begin{equation} \label{eq:source_operator}
  \mathcal{J}(t)
  \equiv
  D_{c'}(f,t) Q_{c'}(f,t),
\end{equation}
where  we employ  $f(\mathbf{x})  \equiv \exp(-|\mathbf{x}|/\rho)$  in this
paper with $\rho$ being the size parameter.

In the large-$t$ region, the four-point correlator is dominated by the ground
state, and the  NBS wavefunction $\psi_{S}(\mathbf{r})$ for the ground state is
obtained as
\begin{equation}
  \psi_{S} (\mathbf{r})
  \propto
  C(\mathbf{r}, t; t_{\rm src})
  \quad
  \text{for large } t.
\end{equation}
Because the  source operator $\mathcal{J}(t)$ transforms  according to the
$A_1^+$ representation  of  the cubic  group,  the extracted  NBS wavefunction
$\psi_{S}(\mathbf{r})$ corresponds to the S-wave.


We define
\begin{equation}
  \widetilde{V}_{0}(\mathbf{r})
  \equiv
  \frac{\nabla^{2} \psi_{S}(\mathbf{r})}{\psi_{S}(\mathbf{r})},
  \label{eq:prepotential}
\end{equation}
which will be referred to as ``prepotential''.
The spatial Laplacian is evaluated using the standard
nearest-neighbour finite-difference discretization,
\begin{equation}
  \nabla^{2}f(\mathbf{r})= \frac{1}{a^{2}}
  \sum_{i=1}^{3}\big[f(\mathbf{r}+a\hat{i}) + f(\mathbf{r}- a\hat{i}) - 2f(\mathbf{r})\big]
\end{equation}
where $a$ is the lattice spacing and $\hat{i}$ denotes a unit vector in the $i$
spatial direction.
By  using the  Schr\"odinger equation~\eqref{eq:schroedinger-eq},  the
prepotential $\widetilde{V}_0(\mathbf{r})$ is related to the potential
$V_0(\mathbf{r})$ through
\begin{equation}
  \widetilde{V}_{0}(\mathbf{r})
  =
  2m_{D}
  \left[
    V_{0}(\mathbf{r}) + m_{D} - \varepsilon_{S}
    \right],
\end{equation}
where $\varepsilon_{S} \equiv \varepsilon_{L = S}$.

We  reexpress   the  Schr\"odinger  equation  by   directly  using  the
prepotential as
\begin{equation}
  \left(
  -\nabla^{2}
  + \widetilde{V}_{0}(r)
  \right)
  \psi_{P}(\mathbf{r})
  =
  \Delta \widetilde{E} \,
  \psi_{P}(\mathbf{r}),
  \label{eq:pre-schroedinger-eq}
\end{equation}
where $\varepsilon_{P} \equiv \varepsilon_{L=P}$ and $\Delta \widetilde{E}
\equiv 2m_D(\varepsilon_P - \varepsilon_S)$.
This equation will be referred to as the ``pre-Schr\"odinger'' equation.

The diquark  mass $m_D$ is  determined by requiring that  the baryonic spectrum
in  the P-wave sector, extracted  from two-point correlators, be      reproduced
by       the      Schr\"odinger      equation, Eq.~\eqref{eq:schroedinger-eq}.
In        practice,        this       amounts        to        solving
Eq.~\eqref{eq:pre-schroedinger-eq}  in  the   P-wave  sector  yielding $\Delta
\widetilde{E}$,
from which the diquark mass is determined as
\begin{equation}
  m_{D}
  =
  \frac{\Delta \widetilde{E}}{2(\varepsilon_P - \varepsilon_S)},
  \label{eq:diquark-mass}
\end{equation}
where $\varepsilon_P$ and  $\varepsilon_S$ are extracted independently from the
corresponding two-point correlators.
The quark-diquark central potential is then obtained through
\begin{equation}
  V_{0}(r)
  =
  \frac{1}{2m_{D}}
  \widetilde{V}_{0}(r)
  - m_{D}
  + \varepsilon_{S}.
  \label{eq:central-potential}
\end{equation}

\section{Lattice Setup}
Our lattice QCD calculations  are performed using the \((2+1)\)-flavor
(\(ud+s\))   gauge    configurations   generated   by    the   PACS-CS
Collaboration~\cite{aoki_2009}.
The ensemble consists of 399 configurations on a lattice of size \(L^{3} \times
T = 32^3 \times 64\), generated with the Iwasaki gauge action at \(\beta = 6/g^2
= 1.9\) and the nonperturbatively $O(a)$-improved Wilson quark action at
\((\kappa_{ud},\kappa_{s}) = (0.13700,0.13640)\) and \(C_{\rm SW} = 1.715\).
This   setup   leads   to   a    lattice   spacing   of   \(a   \simeq
0.0907(13)\,\mathrm{fm}\)               (\(a^{-1}              \approx
2.176(31)\,\mathrm{GeV}\)),  a  spatial lattice  size  of  \(La =  32a \approx
2.902\,\mathrm{fm}\),  a   pion  mass   of  \(m_\pi   \simeq
702(1)\,\mathrm{MeV}\),   and  a   nucleon   mass   of  \(m_N   \simeq
1.583(5)\,\mathrm{GeV}\).

Coulomb gauge fixing is employed before computing the propagators.
The    exponentially-smeared    source    with    smearing    function
$f(r)=\exp(-r/3a)$  is employed  to  enhance ground-state  saturation, where the
smearing  radius was  chosen to  optimize the  ground-state overlap.
To reduce statistical fluctuations, one iteration of
hypercubic (HYP) smearing~\cite{Hasenfratz_Knechtli_2001} with HYP1 parameter
set $[(\alpha_{1},\alpha_{2},\alpha_{3})=(0.75,0.6,0.3)]$ is applied to the
gauge links entering the static-quark propagators (Wilson lines).
%
%
128 source points are employed. They are located at \((N_{x},N_{y},N_{z},N_{t})
= (0,0,0,2n),(8,8,8,2n+1),(16,16,16,2n),(16,16,16,2n+1)\) where
\(n=0,1,2,\dots,31\) to improve the statistical precision.
Further  noise  reduction is  achieved  by  averaging over  the  cubic
rotational  symmetry   of  the  lattice   as  well  as   by  utilizing
time-reversal and charge-conjugation symmetries.
%
For comparison with the quark--diquark potential, we also present the static
$Q\bar{Q}$ potential extracted from Wilson loops, both without HYP smearing and
with the HYP2 parameter set,
$(\alpha_{1},\alpha_{2},\alpha_{3})=(1.0,1.0,0.5)$~\cite{Donnellan_Knechtli_Leder_Sommer_2011}.
For the HYP2 result, one HYP-smearing iteration is applied to both the temporal
and spatial gauge links. To improve the ground-state overlap, APE smearing with
$N_{\mathrm{APE}}=50$ and $\alpha_{\mathrm{APE}}=2.3$ \cite{Takahashi:2000te} is
subsequently applied to the spatial gauge links.
We adopted HYP1 for the quark--diquark calculation following
Ref.~\cite{C_Alexandrou_2006}.  For the Wilson-loop calculation, we tested the
no-HYP, HYP1, and HYP2 prescriptions and found that HYP2 provides the best
statistical precision among the three.  We therefore use the HYP2 result as the
HYP-smeared Wilson-loop result presented below.
Statistical uncertainties are estimated using the jackknife method with a bin
size of 15 configurations unless otherwise indicate.
Unless otherwise stated, the uncertainties quoted for the
quantities extracted in the present analysis are statistical only. The
uncertainty in the lattice scale is not included and corresponds to an
additional common relative uncertainty of approximately $1.4\%$ for quantities
converted to physical units.

\section{Results and Discussion}
\subsection{Effective mass and effective mass differences}
We calculate  the effective masses  of baryonic systems composed  of a static
quark  and  diquarks  in  various  channels,  and  define  the corresponding
effective mass differences by
\begin{equation}
  \Delta m_{C\Gamma} = m_{C\Gamma} - m_{C\gamma_{5}}.
\end{equation}
This gauge-invariant quantity, $\Delta m_{C\Gamma}$ serves as a good estimate of
the mass splitting between two diquark channels, hence exposing the diquark
spectrum.
The effective masses, $m_{C\Gamma}(t) \equiv a^{-1} \log \left[G_{\Gamma}(t) /
G_{\Gamma}(t+a) \right]$ are extracted from the two-point correlators of the
baryonic systems (smeared source and point sink),
\begin{equation}
  G_{\Gamma}(t)
  \equiv
  \left\langle
  J_{\Gamma}(\mathbf{x},t)
  J^\dagger_{\Gamma}(\mathbf{x},0)
  \right\rangle,
\end{equation}
where  
$J_{\Gamma}(x)$ denotes  the interpolating  operator for  the baryonic system
defined  in   Eq.~(\ref{eq:source_operator})  with   $f(x)  = \exp(-r/3a)$ and
$\Gamma = 1,\gamma_\mu, \gamma_5, \gamma_5\gamma_\mu, \sigma_{\mu\nu}$.
The static quark propagator is replaced by a temporal Wilson line:
\begin{equation}
  S_{\rm stat}(\mathbf{x}_2,t_2; \mathbf{x}_1,t_1)
  =
  \delta^3(\mathbf{x}_2 - \mathbf{x}_1)
  \left( \frac{1 + \gamma_0}{2} \right)
  \left[
    \prod_{t=t_1}^{t_2-a} U_4(\mathbf{x}_1,t)
  \right]^\dagger
\end{equation}
for $t_2 >  t_1$, where the factor  $\exp\left(-m_Q(t_2 - t_1)\right)$ is
dropped with $m_Q = \infty$ being the mass of the static quark.

\begin{figure}[htbp]
     \includegraphics[width=0.5\textwidth]{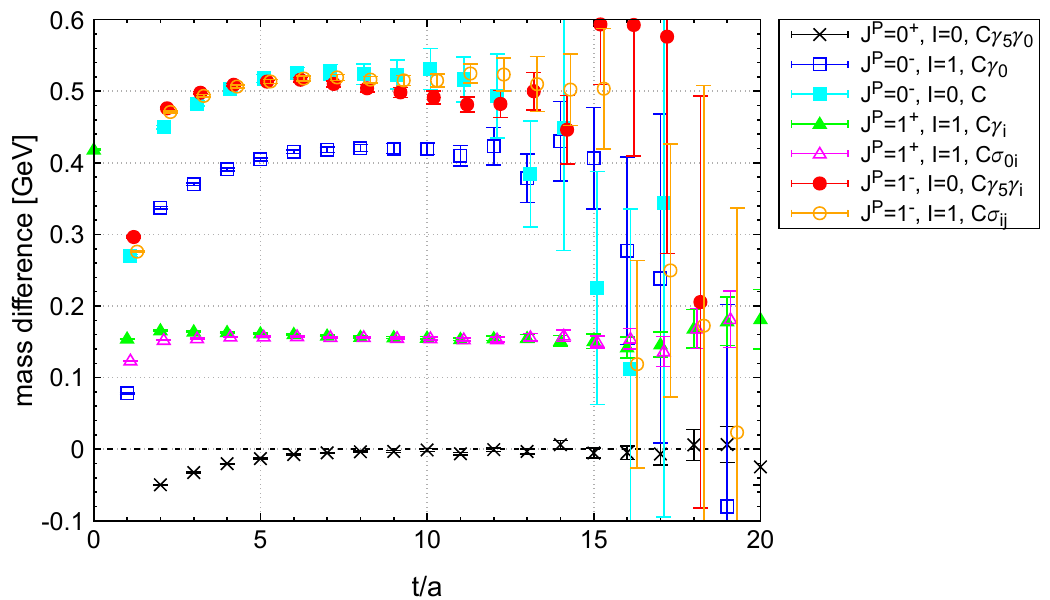}
     \caption{Effective mass difference, $\Delta m_{C\Gamma}$ for various
     $\Gamma$.}
     \label{fig:mass_diff}
\end{figure}
The resulting effective mass differences, $\Delta m_{C\Gamma}$ are shown in
Fig.~\ref{fig:mass_diff}. Effective masses of some baryonic states are also
tabulated in Table~\ref{Table:baryon_mass}.
These masses are extracted by fitting the
two-point correlators to the single-exponential form $C(t)=A\exp(-mt)$ using an
uncorrelated fit. The fit intervals and corresponding $\chi^{2}/d.o.f$ values are
given in Table~\ref{Table:baryon_mass}.

\begin{table}[http]
  \centering
\begin{tabular}{l c c c} 
 \hline
 $J^{P}$ & Mass(GeV) & Fit Range  &$\chi^{2}/d.o.f$       \\ 
 \hline
 $0^{+}$ & 1.585(3)	 & 10  $\leq t \leq$ 19  &0.0208\\
 $1^{+}$ & 1.741(5)	 & 13  $\leq t \leq$ 18  &0.0161\\
 $1^{-}$ & 2.073(12) & 10  $\leq t \leq$ 15  &0.0233\\
 \hline
\end{tabular}
\caption{Effective  mass  for some  baryonic  states  in GeV.  $J^{P}$ indicates
  the total  angular momentum and parity  of the brown-muck, i.e static quark is
  just a spectator.  The fit range used to fit the two-point correlator is also
  listed. The uncertainties in parentheses are statistical only;
  the common lattice-scale uncertainty is not included.}
\label{Table:baryon_mass}
\end{table}

We notice a  tendency that the odd parity states  are heavier than the even
parity states.
In  the  even parity  sector,  the  axial-vector($1^{+}$) channels  are heavier
than the scalar($0^{+}$) channels.
In  the odd  parity sector,  although the  errorbars are  rather large
especially for  $\Gamma = \mathbf{1}$  channel, a similar  tendency is observed,
i.e.,  the  vector($1^-$)  channels  are  heavier  than  the pseudo-scalar
($0^-$) channels.
%
These    results   are    consistent    with    those   reported    in
Ref.~\cite{C_Alexandrou_2006,A_Francis_2022}.

Before proceeding  further, we  comment on the  nature of  the $1^{-}$ state.
This  state can  be interpreted  either as  a $1^{-}$  diquark in  the S-wave
channel or as a $0^{+}$ diquark in the P-wave channel.
In general, the physical state may contain both components.
While  Ref.~\cite{C_Alexandrou_2006,A_Francis_2022} interpreted  this state  as
being dominated  by  the  former  possibility,  we  interpret  it  as  being
predominantly  a  $0^{+}$  diquark  in  the  P-wave  channel,  i.e.  a
$\lambda$-mode excitation.
This     assignment      is     supported      by     phenomenological
studies~\cite{Nagahiro_2017}, and it is crucial for the application of the
quark--diquark  potential framework  used in  the present  work to determine the
mass   of  the  ``good''  scalar   diquark,  i.e.  the $J^{P}=0^{+}$ diquark.

\subsection{Scalar diquark mass, \(m_{D}\) and the quark-diquark potential}
\begin{figure}[htbp]
     \includegraphics[width=0.49\textwidth]{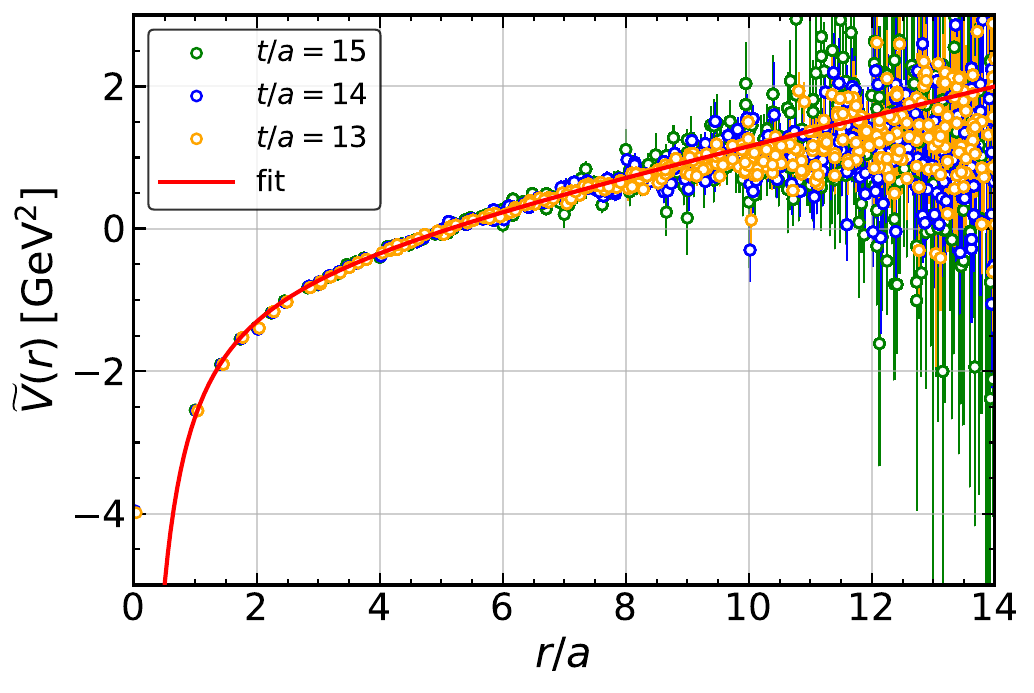}
     \caption{Prepotential at $t/a = 13,14,15$.The red line
     represents the fitted prepotential at $t/a=14$.}
     \label{fig:potential}
\end{figure}

Fig.~\ref{fig:potential} shows the prepotentials obtained from the NBS
wavefunctions at $t/a=13, 14$ and $15$.
We observe that the prepotentials at  $t/a = 14$ and 15 are consistent within
statistical uncertainties, indicating that convergence has been achieved.
We see that these prepotentials are of Cornell-type.
Since the results at $t/a = 14$ and 15 are consistent within errors, we adopt
the prepotential at $t/a=14$ for the subsequent analysis.

The prepotential  is fitted with  the Cornell form within  the fitting range $1
\le r/a \le 10$: 
\begin{equation}
  \tilde{V}(r) = -\frac{\tilde{A}}{r} + \tilde{\sigma} r + \tilde{C}.
  \label{eq:Cornell-form}
\end{equation}
The extracted parameters remain stable when the lower bound is
varied to $r/a=2$ and $3$. By using  this fitted form,  we solve the
pre-Schr\"odinger equation, Eq.~(\ref{eq:pre-schroedinger-eq}), where the
shooting method is used with the Dormand-Prince fifth-order Runge-Kutta (RK)
algorithm.
For the  ground state,  we obtain $\Delta  \widetilde{E} \simeq 0.911$ GeV$^2$.
It leads to
\begin{equation}
  m_D
  =
  \frac{\Delta   \widetilde{E}}{2(\varepsilon_P   -  \varepsilon_S)}
  =
  0.931(7)\,\,\mbox{GeV},
\end{equation}
which  is  close  to  the naive  constituent-quark  estimate,  $2m_N/3 \approx
1.05~\mathrm{GeV}$.

\begin{figure}[htbp]
     \includegraphics[width=0.49\textwidth]{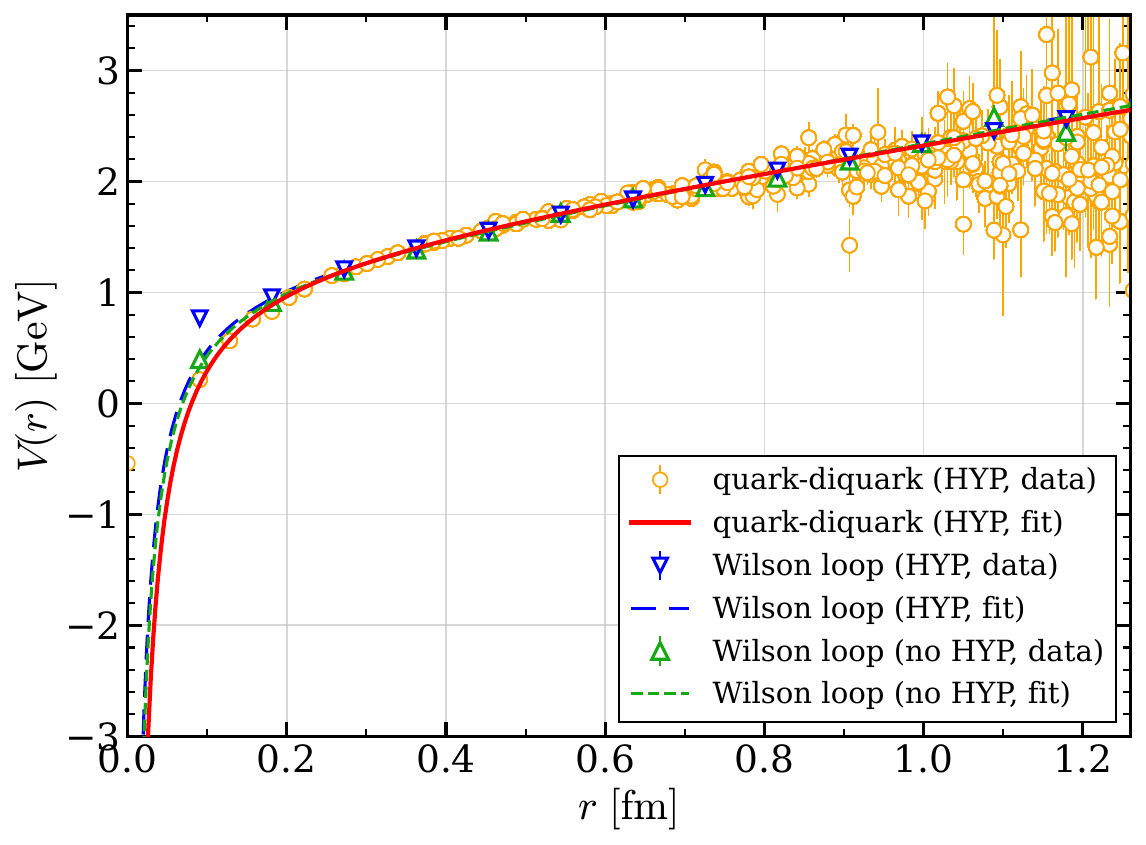}
     \caption{The quark–diquark potential at $t/a = 14$ together with the
       $Q\bar{Q}$ potential extracted from the Wilson loop.
       The  quark-diquark   data  (circles)  with  its   Cornell  fit (red line)
       is compared  against the  $Q\bar{Q}$ data (triangles  for without  HYP,
       inverted triangles for with  HYP) with their Cornell fit (green and blue
       dashed lines).
       To ease  the comparison, the $Q\bar{Q}$  potentials are shifted
       vertically. 
       The HYP-smeared quark-diquark and Wilson-loop
       calculations use the HYP1 and HYP2 parameter sets, respectively.}
     \label{fig:qqQ_vs_QQbar_potential}
\end{figure}

The     resulting    quark-diquark     potential    is     shown    in
Fig.~\ref{fig:qqQ_vs_QQbar_potential}.  It is  fitted with the Cornell form
within the fitting range $a \le r \le 10a$:
\begin{equation}
  V(r) = -\frac{A}{r} + \sigma r + C.
\end{equation}
%
The results are given in Table~\ref{Table:parameter_potential}.
%
%
%
\begin{table}[t]
\centering
\begin{tabular}{ccc}
\hline
Potential & $A$ (GeV$\cdot$fm) & $\sqrt{\sigma}$ (GeV) \\
\hline
Quark--diquark, HYP       & 0.110(3) & 0.475(6) \\
Wilson loop, HYP         &0.082(2) & 0.498(3) \\
Wilson loop, without HYP  & 0.087(3) & 0.497(3) \\
\hline
\end{tabular}
\caption{The Coulomb coefficient $A$ and string tension
$\sqrt{\sigma}$ extracted from the quark--diquark potential and the static
$Q\bar{Q}$ potentials obtained from Wilson loops. The HYP-smeared quark--diquark
and Wilson-loop calculations use the HYP1 and HYP2 parameter sets,
respectively.}
\label{Table:parameter_potential}
\end{table}
%
%
%
The  uncertainties  of  the  fitted parameters  are  estimated  by  an
approximate  double-nested  jackknife   procedure.   Details  will  be reported
elsewhere.

Figure~\ref{fig:qqQ_vs_QQbar_potential}  compares  the  quark–diquark potential
with   the  static  $Q\bar{Q}$  potentials.    (The  static $Q\bar{Q}$
potentials are vertically shifted to ease the comparison.)
The static  $Q\bar{Q}$ potentials  are fitted  with the  Cornell form,
Eq.(\ref{eq:Cornell-form}),  within the  fitting range  $3a \le  r \le 16a$  and
the  resulting  fit parameters  for  the HYP-smeared  static $Q\bar{Q}$
potential         are         summarized         in
Table~\ref{Table:parameter_potential}.
%
The  agreement between  the  quark--diquark potential  and the  static
$Q\bar{Q}$ potential is remarkably good.
In   particular,   the   extracted  string   tensions   agree   within
approximately  5\%,   indicating  that  the  confining   part  of  the
quark–diquark  potential  is  compatible  with  that  of  the  static $Q\bar{Q}$
potential.
However, the short-distance  behavior ($r \lesssim 2a  \simeq 0.2$ fm) should be
interpreted with caution,  since HYP smearing  affects this region.
Indeed, the  HYP-smeared static $Q\bar{Q}$ potential  shows noticeable
deviations from the Cornell form at short distances.

The Schr\"odinger equation, Eq.~(\ref{eq:schroedinger-eq}), allows the total
baryon  energy to be  decomposed into  the diquark mass  and the remaining
energy contribution,
\begin{equation}
  m_B = m_D + E_{\rm rem}.
\end{equation}
For the S-wave baryon state, we obtain
\begin{equation}
  1.585~{\rm GeV}
  =
  0.931~{\rm GeV}
  +
  0.654~{\rm GeV},
\end{equation}
where the first term corresponds to the scalar diquark mass, while the second
term represents the remaining contribution to the baryon energy.
It  should  be  noted  that  this  remaining  contribution  cannot  be
interpreted directly as the quark--diquark interaction energy.
In general,  it contains  both the quark--diquark  interaction energy, $E_{\rm
Int}$,  and the  self-energy  of  the static  quark,  $E_{\rm self}$:
\begin{equation}
  E_{\rm rem}
  =
  E_{\rm Int}
  +
  E_{\rm self}.
\end{equation}
The static-quark self-energy is ultraviolet divergent in the continuum limit, $a
\to 0$.
Note that a further separation of $E_{\rm rem}$ into $E_{\rm Int}$ and $E_{\rm
self}$  is scheme  dependent and  will not  be pursued  in the present work.

Nevertheless,  the static-quark  self-energy is  common to  all baryon channels
and therefore cancels in mass differences.
For example,
\begin{equation}
  m_B(1^+) - m_B(0^+)
  =
  m_D(1^+) - m_D(0^+)
  +
  \Delta E_{\rm Int}.
\end{equation}
This relation clarifies that  baryon mass splittings generally receive
contributions  not only  from diquark  mass splittings  but also  from
differences in the quark--diquark interaction energy.

The present value, $m_D = 0.931(7)$ GeV, is substantially smaller than the value
reported in Ref.~\cite{watanabe_2021, K_watanabe_2022}.
Interestingly,   our   result   is    much   closer   to   the   naive
constituent-quark estimate,  $2m_N/3 \simeq 1.05$ GeV.
The   origin  of   the   discrepancy  with   Ref.~\cite{watanabe_2021,
  K_watanabe_2022}  has  been  clarified   through  a  more  extensive analysis
  and will be reported elsewhere.

\section{Summary and conclusions}

We performed a lattice QCD  investigation of baryonic systems composed of a
diquark and  a static quark with the aim  of determining both the diquark   mass
and   the  corresponding   quark–diquark  interaction potential.
Our analysis was  carried out using $2+1$  flavor gauge configurations generated
by  the PACS-CS collaborations  on a lattice of  size $L^{3} \times T  = 32^{3}
\times 64$  for the pion mass  $m_{\pi} \simeq 702$ MeV.

We  first evaluated  the effective  mass differences  between baryonic states
containing different  diquark  operators and  a static  quark, following the
strategy employed in Ref.~\cite{C_Alexandrou_2006,A_Francis_2022}.
By neglecting the interaction energy  between the static quark and the diquark,
these  differences  can be  interpreted  as  gauge-invariant estimates of the
corresponding diquark mass splittings.
Our    results    are    consistent    with    those    reported    in
Ref.~\cite{C_Alexandrou_2006},  indicating   that  the   present  setup
successfully   reproduces  the   previously   observed  diquark   mass
hierarchy.

In order to  obtain the diquark mass without relying  on the existence of the
bound-state poles in the diquark two-point correlators,
we adopted a  quark–diquark description in which the  diquark mass is treated as
an effective parameter.
The quark–diquark  interaction potential  was determined using  a HAL QCD-based
method applied  to baryon-like systems composed  of a scalar (``good'') diquark
and a static quark.
Employing a static  color source eliminates the  uncertainty caused by the
choice  of   the  charm  quark  mass  that   appears  in  earlier quark–diquark
studies~\cite{watanabe_2021,      K_watanabe_2022, Nishioka_2025,
Nishioka:2026oca}.
Following the  same strategy  as employed  in Ref.~\cite{watanabe_2021,
  K_watanabe_2022},  the diquark  mass  was determined  by imposing  a
  consistency condition  between two  independent determinations  of the
  baryonic energy in  P-wave channel: one obtained from  the solution of the
  Schrödinger  equation and  the  other  derived from  conventional two-point
  correlation functions.
Using this procedure, we obtain a scalar diquark mass $m_D = 0.931(7)$ GeV.
This value is  close to a na\"ive constituent  quark estimate, $2m_N/3 \simeq
1.05$ GeV.

As an important advantage, our framework allows the baryon-like energy to  be
decomposed  into  the  diquark mass  and  the remaining  energy contribution.
This  separation provides  a  more transparent  interpretation of  the extracted
diquark  mass than  in  previous  studies based  solely  on baryon-like energy
splittings.

We  also  obtained  the  quark–diquark potential,  which  exhibits  a
Cornell-type (Coulomb+linear) form.
The resulting string tension, $\sqrt{\sigma} = 475(6)$ MeV, agrees with the
static quark-antiquark ($Q\bar{Q}$) potential extracted from Wilson loop
calculation.

Several directions for future investigation remain.
First, the present framework will be extended to study the axial-vector diquark,
$J^{P} = 1^{+}$ channel.
With this, the mass splittings between $1^{+}$ and $0^{+}$ diquarks can be
systematically studied under this framework.
Then, calculations at multiple quark masses will be carried out to examine the
quark-mass dependence of diquark properties.

\section*{Acknowledgements}
Numerical   calculations  in   this   work  were   performed  on   the
supercomputer SQUID at  D3 Center, The University  of Osaka, supported by the
Research  Center for Nuclear Physics (RCNP),  The University of Osaka.
This  work was  (partly)  achieved through  the  use of  Supercomputer System
SQUID at  the D3 Center, The University of  Osaka under Project for Nurturing
Student Competing with the World.
We  would like  to  thank the  PACS-CS Collaboration  as  well as  the JLDG/ILDG
for providing the $2+1$ flavor QCD gauge configurations.
We employed a modified version of  the lattice QCD library Bridge++ to carry out
our computations \cite{Ueda_2014}.
This  work was  supported by  JSPS KAKENHI  Grant Numbers  JP21K03535,
JP25K07323 and JP23H05439.
We  also  acknowledge the  support  from  the Ministry  of  Education, Culture,
Sports,  Science and Technology  (MEXT) of Japan  through the Japanese
Government Scholarship.

\appendix



\bibliographystyle{elsarticle-num} 
\bibliography{ref}

@article{Jaffe_2005,
  author    = {R. L. Jaffe},
  title     = {Exotica},
  journal   = {Nucl. Phys. B Proc. Suppl.},
  volume    = {142},
  year      = {2005},
  pages     = {343}
}

@article{C_Alexandrou_2006,
  author    = {C. Alexandrou and Ph. de Forcrand and B. Lucini},
  title     = {Evidence for diquarks in lattice QCD},
  journal   = {Phys. Rev. Lett.},
  volume    = {97},
  year      = {2006},
  pages     = {222002},
  eprint    = {hep-lat/0609004}
}

@article{A_Francis_2022,
  author    = {A. Francis and P. de Forcrand and R. Lewis and K. Maltman},
  title     = {Diquark properties from full QCD lattice simulations},
  journal   = {JHEP},
  volume    = {05},
  year      = {2022},
  pages     = {062},
  eprint    = {2106.09080}
}

@article{K_watanabe_2022,
  author    = {Kai Watanabe},
  title     = {Quark-diquark potential and diquark mass from lattice QCD},
  journal   = {Phys. Rev. D},
  volume    = {105},
  year      = {2022},
  pages     = {074510},
  eprint    = {2111.15167}
}

@inproceedings{Orginos_2006,
  author    = {K. Orginos},
  title     = {Diquark properties from lattice QCD},
  booktitle = {PoS LAT2005},
  year      = {2006},
  pages     = {054},
  eprint    = {hep-lat/0510082}
}

@inproceedings{Green_2010,
  author    = {J. Green and J. Negele and M. Engelhardt and P. Varilly},
  title     = {Spatial diquark correlations in a hadron},
  booktitle = {PoS LATTICE2010},
  year      = {2010},
  pages     = {140},
  eprint    = {1012.2353}
}

@article{Hess_1997,
  author    = {M. Hess and F. Karsch and E. Laermann and I. Wetzorke},
  title     = {Diquark masses from lattice QCD},
  journal   = {Phys. Rev. D},
  volume    = {58},
  year      = {1997},
  pages     = {111502},
  eprint    = {hep-lat/9804023}
}

@article{Babich_2007,
  author    = {R. Babich and N. Garron and C. Hoelbling and J. Howard and L. Lellouch and C. Rebbi},
  title     = {Diquark correlations in baryons on the lattice with overlap quarks},
  journal   = {Phys. Rev. D},
  volume    = {76},
  year      = {2007},
  pages     = {074021},
  eprint    = {hep-lat/0701023}
}

@article{Bi_2016,
  author    = {Y. Bi and Hao Cai and Y. Chen and M. Gong and Z. Liu and H.-X. Qiao and Y.-B. Yang},
  title     = {Diquark mass differences from unquenched lattice QCD},
  journal   = {Chin. Phys. C},
  volume    = {40},
  year      = {2016},
  pages     = {073106},
  eprint    = {1510.07354}
}

@article{Nishioka_2025,
  author    = {S. Nishioka and N. Ishii},
  title     = {Axialvector diquark Mass and quark-diquark potential in \({\Sigma}_{c}\)},
  journal   = {PoS (LATTICE2024)},
  volume    = {305},
  year      = {2025},
  eprint    = {2502.06260 }
}

@article{Nishioka:2026oca,
    author = "Nishioka, Soya and Ishii, Noriyoshi",
    title = "{Axialvector diquark Mass and quark-diquark potential in $\Sigma_c$}",
    doi = "10.22323/1.500.0013",
    journal = "PoS",
    volume = "HADRON2025",
    pages = "013",
    year = "2026"
}

@article{Shuryak_1982, 
    title={The role of Instantons in Quantum Chromodynamics}, volume={203}, 
    DOI={10.1016/0550-3213(82)90478-3}, 
    number={1}, 
    journal={Nuclear Physics B},
    author={Shuryak, E.V.}, 
    year={1982}, 
    pages={93–115}
}

@article{Takahashi:2000te,
    author = "Takahashi, Toru T. and Matsufuru, H. and Nemoto, Y. and Suganuma, H.",
    title = "{The Three quark potential in the SU(3) lattice QCD}",
    eprint = "hep-lat/0006005",
    archivePrefix = "arXiv",
    doi = "10.1103/PhysRevLett.86.18",
    journal = "Phys. Rev. Lett.",
    volume = "86",
    pages = "18--21",
    year = "2001"
}

@article{Shuryak_Zahed_2004,
    title={A schematic model for pentaquarks based on Diquarks}, 
    volume={589}, 
    DOI={10.1016/j.physletb.2004.03.019}, 
    number={1–2}, 
    journal={Physics Letters B}, 
    author={Shuryak, Edward and Zahed, Ismail}, 
    year={2004}, 
    pages={21–27}
}

@article{tHooft_1976, 
    title={  
    }, 
    volume={14}, 
    DOI={10.1103/physrevd.14.3432}, 
    number={12}, 
    journal={Physical Review D}, 
    author={’t Hooft, G.}, 
    year={1976}, 
    pages={3432–3450}
}

@article{Schäfer_Shuryak_1998, 
    title={Instantons in QCD}, 
    volume={70}, 
    DOI={10.1103/revmodphys.70.323}, 
    number={2}, 
    ournal={Reviews of Modern Physics},
    author={Schäfer, T. and Shuryak, E. V.}, 
    year={1998}, 
    pages={323–425}
}

@article{Nagahiro_2017, 
    title={Structure of charmed baryons studied by pionic decays}, 
    volume={95}, 
    DOI={10.1103/physrevd.95.014023}, 
    number={1}, 
    journal={Physical Review D}, 
    author={Nagahiro, Hideko and Yasui, Shigehiro and Hosaka, Atsushi and Oka, Makoto and Noumi, Hiroyuki}, 
    year={2017}
}

@article{Workman_2022, 
    title={Review of Particle Physics}, 
    volume={2022}, 
    DOI={https://doi.org/10.1093/ptep/ptac097}, 
    number={8}, 
    journal={Progress of Theoretical and Experimental Physics}, 
    author={R.L. Workman and others}, 
    year={2022}, 
    pages={083C01}
}

@article{Hasenfratz_Knechtli_2001, 
    title={Flavor symmetry and the static potential with hypercubic blocking}, 
    volume={64},
    DOI={10.1103/physrevd.64.034504}, 
    number={3},
    journal={Physical Review D},
    author={Hasenfratz, Anna and Knechtli, Francesco},
    year={2001}
}

@article{Ueda_2014,
    author = {S. Ueda and S. Aoki and T. Aoyama and K. Kanaya and H. Matsufuru and S. Motoki and Y. Namekawa and H. Nemura and Y. Taniguchi and N. Ukita},
    title = {Development of an object oriented lattice QCD code ’Bridge++’},
    journal = {J. Phys. Conf. Ser. 523 012046.},
    number = {523},
    year = {2014}
}

@article{watanabe_2021,
    author = {K. Watanabe and N. Ishii},
    title = {Building Diquark Model from Lattice QCD},
    journal = {Few Body Syst.},
    year = {2021},
    pages = {45},
    volume = {23},
    number = {3}
}

@article{aoki_2009,
    author = {S. Aoki and K.-I. Ishikawa and N. Ishizuka and T. Izubuchi and D. Kadoh and K. Kanaya and Y. Kuramashi and Y. Namekawa and M. Okawa and Y. Taniguchi and A. Ukawa and N. Ukita and T. Yoshie},
    title = {2 + 1 flavor lattice QCD toward the physical point},
    journal = {Phys. Rev. D},
    year = {2009},
    pages = {034503},
    volume = {79}
}

@article{Donnellan_Knechtli_Leder_Sommer_2011,
    title={Determination of the static potential with dynamical fermions},
    volume={849},
    DOI={10.1016/j.nuclphysb.2011.03.013},
    number={1},
    journal={Nuclear Physics B},
    author={Donnellan, Michael and Knechtli, Francesco and Leder, Björn and Sommer, Rainer},
    year={2011},
    pages={45–63}}





\end{document}